\documentclass[pdflatex,showpacs,superscriptaddress,preprint,nofootinbib]{revtex4-1}
\usepackage{bm,amsmath,amssymb}
\usepackage{graphicx}
\usepackage{subfigure}
\usepackage{color}
\usepackage{soul}

\usepackage{multirow}
\usepackage{lineno}
\usepackage{footnote}
\usepackage{algpseudocode,algorithm}
\RequirePackage{xspace}
\usepackage{mathrsfs}
\usepackage{hyperref}
\usepackage{bm}
\usepackage{xcolor}

\newcommand{\fig}{Fig.\xspace}
\newcommand{\tab}{Tab.\xspace}

\newcommand{\suthree}{SU(3)\xspace}
\newcommand{\uone}{U(1)\xspace}
\newcommand{\Tr}{\mathrm{Tr}\,}  %capital Trace
\newcommand{\av}[1]{\left\langle #1 \right\rangle} %Def of <average>
\renewcommand{\bar}[1]{\overline{#1}}%
\if0{%-- For exam.
\newcounter{nombre} %Mandatory
\setcounter{nombre}{0} %If it necessary
\newcounter{nombresub}[nombre] %define hierarchy
\setcounter{nombresub}{0} %If it necessary

}\fi
\begin{document}
%	\linenumbers

%
% Main-part !!
%
%

\title{Chiral phase structure of three flavor QCD \\in a background magnetic field }
\preprint{\today}
\author{Heng-Tong Ding}
\email[]{hengtong.ding@mail.ccnu.edu.cn}
\affiliation{Key Laboratory of Quark \& Lepton Physics (MOE) and Institute of Particle Physics, Central China Normal University, Wuhan 430079, China}
\author{{Christian Schmidt} }
\email[]{schmidt@physik.uni-bielefeld.de}
\affiliation{Fakult\"at f\"ur Physik, Universit\"at Bielefeld, D-33615 Bielefeld, Germany}        
\author{Akio Tomiya}
\email[]{akio.tomiya@riken.jp}
\affiliation{RIKEN/BNL Research center, Brookhaven National Laboratory, 
Upton, NY, 11973, USA}
\author{Xiao-Dan Wang}
\email[]{xiaodanwang@mails.ccnu.edu.cn}
\affiliation{Key Laboratory of Quark \& Lepton Physics (MOE) and Institute of Particle Physics, Central China Normal University, Wuhan 430079, China}

\begin{abstract}
We investigate the chiral phase structure of
three flavor QCD in a background $U(1)$ magnetic field using
the standard staggered action and the Wilson plaquette gauge action.
We perform simulations on lattices with a temporal extent of $N_\tau=4$ and four spatial extents of $N_\sigma = 8,16, 20$ and 24. We choose a quark mass in lattice spacing as $am = 0.030$ with corresponding pion mass estimated as $m_\pi\sim 280$ MeV such that there exists a crossover transition at vanishing magnetic fields, and adopt two values of magnetic field strength in lattice spacing $a \sqrt{ e{B}}\simeq 1.5$ and 2 corresponding to $eB/m_\pi^2\sim11$ and 20, respectively. We find that the transition becomes stronger in the presence of a background magnetic field, and turns into a first order as seen from the volume scaling of the order parameter susceptibility as well as the metastable states in the time history of the chiral condensate. On the other hand, the chiral condensate and transition temperature always increase with $B$ even within the regime of a first order phase transition. This suggests that the discrepancy in the behavior of chiral condensates and transition temperature as a function of $B$ between earlier lattice studies using larger-than-physical pion masses with standard staggered fermions and those using physical pions with improved staggered fermions is mainly due to lattice cutoff effects.
\end{abstract}
%\pacs{}
%\keywords{}

\maketitle

\clearpage
\newcommand{\bhat}{\hat{b} }
\section{Introduction}
Strong magnetic fields have been shown to have many significant impacts on the properties of systems governed by strong interaction, and they may have observable consequences in heavy-ion collision experiments as well as magnetized neutron stars~\cite{Kharzeev:2012ph,Miransky:2015ava}. One of the interesting impacts is the behavior of transition temperature of strong-interaction system and chiral condensate as a function of the magnetic field strength $B$. Early lattice studies of $N_f=2$ QCD with standard staggered fermions and larger-than-physical pions on $N_\tau=4$ lattices found the so-called magnetic catalyses, which means that the chiral condensate increases monotonically with increasing $B$~\cite{DElia:2010abb}. As the chiral symmetry breaking is enhanced it is expected and observed that the transition temperature consequently increases with $B$~\cite{DElia:2010abb}. However, based on continuum-extrapolated lattice results of $N_f=2+1$ QCD using improved staggered fermions and physical pions~\cite{Bali:2011qj} it turns out that the transition temperature has the opposite behavior in magnetic field as compared to earlier studies in Ref.~\cite{DElia:2010abb}. I.e. the transition temperature decreases with increasing $B$. The chiral condensate, on the other hand,  first increases and then decreases with $B$ in the transition regime~\cite{Bali:2011qj}. This non-monotonic behavior of chiral condensate in $B$, which is called inverse magnetic catalysis, has also been observed in further lattice studies~\cite{Ilgenfritz:2013ara,Bornyakov:2013eya,Bali:2014kia}. 

The discrepancy of lattice results in the behavior of chiral condensates and transition temperature in $B$ reported in Refs~\cite{DElia:2010abb} and~\cite{Bali:2011qj,Ilgenfritz:2013ara,Bornyakov:2013eya,Bali:2014kia} is probably due to either large quark masses or possible large cutoff effects present in the first study. To understand the role of quark masses in the intricate relation between the (inverse) magnetic catalysis and the reduction of transition temperature, authors of Refs.~\cite{DElia:2010abb} and~\cite{Bali:2011qj} have performed lattice studies of $N_f$=2+1~\cite{DElia:2018xwo} and $N_f$=3 QCD ~\cite{Endrodi:2019zrl} using improved staggered fermions with various larger-than-physical values of pions (370 MeV $\lesssim m_\pi\lesssim$ 700 MeV) on $N_\tau$=6 lattices. It is found that the reduction of transition temperature always holds, however, the inverse magnetic catalysis disappears at a certain value of pion mass. It is suggested in Ref.~\cite{DElia:2018xwo}  that the inverse magnetic catalysis is more like a deconfinement catalysis~\cite{Bonati:2016kxj} as it is not necessarily associated with the reduction of the transition temperature as a function of $B$.

Despite the discrepancy mentioned above the strength of the QCD transition always becomes larger in the stronger magnetic field as presented in Refs.~\cite{DElia:2010abb,Endrodi:2015oba}. And it is speculated that the strength could be sufficiently large to turn the crossover transition into a first order phase transition which can then generate a critical point in the QCD phase diagram in the plane of temperature and magnetic field~\cite{Endrodi:2015oba,Cohen:2013zja}. However, no true phase transition of QCD in a background magnetic field has been observed in lattice QCD simulations. 

One of the main motivations of the current study is to
explore the chiral phase structure of $N_f=3$ QCD in a background magnetic field. At the vanishing magnetic field the true first order phase transition is not yet observed, and state-of-the-art estimates on the critical pion mass $m_\pi^c$ based on lattice QCD simulations are $m_\pi^c\lesssim50$ MeV using improved staggered fermions~\cite{Endrodi:2007gc,Bazavov:2017xul} and $m_\pi^c\lesssim110$ MeV using clover-improved Wilson fermions~\cite{Nakamura:2019gyy,Kuramashi:2020meg}. Since the background magnetic field always enhances the strength of the transition one may wonder whether it could enlarge the first order chiral phase transition region in $N_f$=3 QCD, i.e. having a larger value of the critical pion mass. 

 In this paper we investigate the transition of $N_f=3$ QCD in background magnetic fields with a quark mass corresponding to pion mass estimated as $\sim280$ MeV at vanishing magnetic field. In our lattice simulations we use standard staggered fermions for a testbed towards to probe a first-order phase transition with magnetic fields using improved fermions, e.g. Highly Improved Staggered Quarks~\cite{Tomiya:2019nym}. The usage of standard staggered fermions with a small quark mass also renders us to understand whether the discrepancy in the behavior of chiral condensate and transition temperature as a function of the magnetic field strength in~\cite{DElia:2010abb,Endrodi:2015oba} is ascribed to the lattice cutoff effects.
 
 The paper is organized as follows.
In Section \ref{sec:setup}, we introduce our lattice setup and observables to be investigated.
In Section \ref{sec:results},  we mainly discuss the order of the transition based on results of chiral condensates, Polyakov loops as well as their susceptibilities and Binder cumulants.
In Section \ref{sec:conclusion},  we conclude our work. The preliminary results have been reported in~\cite{Tomiya:2017cey}.

\section{Lattice setup  and observables} \label{sec:setup}
We perform our simulations on $N_\sigma^3\times N_\tau$ lattices with 3 mass-degenerate flavors of standard staggered quarks and the Wilson plaquette gauge action by employing the rational hybrid Monte Carlo algorithm \cite{Clark:2006wq}. Simulations are performed on lattices with a fixed value of temporal extent $N_\tau=4$ and four different values of spatial size $N_\sigma = 8, 16, 20$ and 24.  Since the critical quark mass, where the first order chiral phase transition starts for this lattice setup is $am_c=$0.027~\cite{Smith:2011pm}, we choose the value of quark mass in lattice spacing $am$ to be 0.03 in our simulations such that the QCD transition is a crossover at vanishing magnetic field.  The temperature, $T=1/(a N_\tau)$ is varied through the relation between the lattice spacing $a$ and the inverse gauge coupling $\beta$, and specifically temperature increases with the value of $\beta$. The background $U(1)$ magnetic field is implemented in a conventional way \cite{AlHashimi:2008hr} and will be reviewed in the following subsection~\ref{sec:mag}. The relevant observables will be introduced in subsection~\ref{sec:obs}.

\subsection{Magnetic fields on the lattice} \label{sec:mag}
Magnetic fields couple to quarks with their electric charges and through covariant derivative in the continuum,
\begin{align}
D_\mu = \partial_\mu - ig A_\mu + iq a_\mu,
\end{align}
where $g$ is the \suthree gauge coupling and $q$ is the electric charge of a quark.
$A_\mu$ and $a_\mu$ are gauge fields for \suthree and \uone, respectively.
On the lattice the background $U(1)$ magnetic field is introduced by substituting the \suthree  link $U_\mu$ by its product with the \uone link $u_\mu$ in the lattice Dirac operator,
\begin{align}
D[U] \to D[uU]. 
\end{align}
In our simulations the magnetic field only points along the $z$-direction.
Since the $x$--$y$ plane has boundaries for a finite system size,
appropriate boundary conditions need to be imposed.
Besides, the magnetic field is realized as a \uone plaquette, and
it introduces non-trivial conditions to the magnitude of the magnetic field as will be depicted next.

Let us denote the lattice size $(N_x,\; N_y,\; N_z,\; N_t)$ and coordinate as $n_\mu=0,\cdots, N_\mu-1$ ($\mu = x,\; y,\; z,\; t$).
The background magnetic field pointing along the $z$-direction $\vec{B}=(0,0,B)$ is described by the link variable $u_\mu(n)$ of the \uone field, and 
$u_\mu(n)$ is expressed as follows in the Landau gauge \cite{AlHashimi:2008hr,Bali:2011qj},
\begin{align}
u_x(n_x,n_y,n_z,n_t)&=
\begin{cases}
\exp[-iq a^2 B N_x n_y] \;\;&(n_x= N_x-1)\\
1 \;\;&(\text{otherwise})\\
\end{cases}\notag\\
u_y(n_x,n_y,n_z,n_t)&=\exp[iq a^2 B  n_x],\label{eq:def_mag_u} \\
u_z(n_x,n_y,n_z,n_t)&=u_t(n_x,n_y,n_z,n_t)=1.\notag
\end{align}
The periodic boundary condition for \uone links is applied for all directions except for the $x$-direction.
One-valuedness of the one-particle wave function along with a plaquette requires the following quantization, 
\begin{align}
a^2 q B = \frac{2\pi N_b}{N_x N_y} \label{eq:mag_to_Nb},
\end{align}
where $N_b \in { \boldsymbol Z}$ is the number of magnetic flux through the unit area for the $x$-$y$ plane.
The ultraviolet cutoff $a$ also introduces a periodicity of the magnetic field along with $N_b$.
Namely, a range
$0\leq N_b < {N_x N_y}/{4}$, 
represents an independent magnitude of the magnetic field $B$.
In our simulations the 3 mass-degenerate flavors are of up, down and strange quark type, and thus two of the flavors have electrical charge of $q_{d,s}=-\frac{e}{3}$ and the rest one has $q_u=\frac{2e}{3}$ with $e$ the electric charge of electron. Thus to satisfy the quantization condition Eq.~\ref{eq:mag_to_Nb} we take the electric charge $q$ to be that of down quark type, i.e. $q=q_{d,s}=-\frac{e}{3}$.
To simplify the notation we use $\hat{b}$ expressed as follows to denote the magnetic field strength
\begin{align}
\bhat \equiv a\sqrt{eB} = \sqrt{ \frac{6\pi N_b}{N_x N_y} }\,\, . \label{eq:magnetic_flux}
\end{align}
 We choose certain values of $N_b$ such that $\hat{b}$ are the same in physical units among various lattice sizes which are listed in Table~\ref{tab:Nb}.

\begin{table}[htb]
\begin{minipage}{0.2\hsize}
\begin{tabular}{c||c|c|c|c|c|c|}
$N_b$& 0& 8  \\ \hline
$\hat{b}$ & 0& 1.5 
\end{tabular}\\
{$N_\sigma=8$}
\end{minipage}
%\hspace{5mm}
\begin{minipage}{0.2\hsize}
\begin{tabular}{c||c|c|c|c|c|c|}
$N_b$& 0& 32 &56 \\ \hline
$\hat{b}$& 0.0& 1.5 &2.0
\end{tabular}\\
{$N_\sigma=16$}
\end{minipage}
\begin{minipage}{0.2\hsize}
\begin{tabular}{c||c|c|c|c|c|c|}
$N_b$& 0 & 50\\ \hline
$\hat{b}$& 0.0 &1.5
\end{tabular}\\
{$N_\sigma=20$}
\end{minipage}
%\hspace{5mm}
\begin{minipage}{0.2\hsize}
\begin{tabular}{c||c|c|c} 
$N_b$& 0& 72\\ \hline
$\hat{b}$& 0.0& 1.5
\end{tabular}\\
{$N_\sigma=24$}
\end{minipage}
\caption{Number of magnetic flux $N_b$ used in our simulations and the approximated values of $\hat{b}\equiv a\sqrt{eB}$ for $N_\sigma=8$,  16, 20 and 24. 
}
\label{tab:Nb}
\end{table}
Based on an earlier estimate of the pion mass in $N_f=3$ QCD~\cite{Karsch:2001nf} we make a estimate of the scale at the pseudo-critical temperature $T_{pc}^{0}:=T_{pc}(\hat{b}=0)$. We find $m_\pi\sim 280$~MeV and $m_\pi/T_{pc}^0\simeq1.77$ at $\hat{b}=0$. The values of magnetic field strength for $\hat{b}=1.5$ and 2 thus correspond to $eB/(T_{pc}^0)^2\simeq36$ and 64, i.e. $eB \sim 11\; m_\pi^2$ and $eB \sim 20\; m_\pi^2$, respectively. We further note that the temperature in the explored $\beta$-intervall [5.13,5.19] may change by approximately $25$~MeV, which would lead to corresponding changes in the magnetic field.

Our simulation parameters and statistics are summarized in 
\tab \ref{tab:stat_L8} - \ref{tab:stat_L24}. 
It is worth mentioning that for our largest lattices, i.e. with $N_\sigma=24$, we have performed simulations with a hot start (configuration with random elements) and a cold start (configuration with unit elements) to check metastable states around the transition temperature to overcome small tunnelling rates.

\subsection{Observables}
\label{sec:obs}
An expectation value for an operator ${O}$ for given gauge coupling $\beta$, mass and magnetic flux ${N_b}$ is defined by,
\begin{align}
\av{O}_{\beta,m,N_b}
&= \int \mathcal{D} U P_{\beta,m,N_b}[U] \; {O}[U],
\label{eq:O} 
\end{align}
\begin{align}
P_{\beta,m,N_b}[U]&= \frac{1}{Z_{\beta,m,N_b}}
\det\big[D_\text{2/3}[U,N_b] + m\big]^{1/4}
\det\big[D_\text{-1/3}[U,N_b] + m\big]^{2/4}
e^{-S_\text{gauge}[U;\beta]},
\end{align}
where  $D$ is the standard staggered Dirac operator and its
subscript indicates the electric charge of the related quark. 
The fractional power, e.g. 1/4 and 2/4,  is due to the fourth root of staggered fermions in our simulations.

We measure the chiral condensate for a down type quark~\footnote{
In principle, we can average up, down and strange quark condensate,
or up and down. However, we choose down condensate to avoid such arbitrariness.},
\begin{align}
 \av{\bar{\psi}\psi} 
= \frac{1}{4N_\sigma^3 N_\tau}
\av{ \Tr\frac{1}{D_{-1/3}+m}  } ,
\end{align}
where $\Tr[\cdots]$ is a trace over color and space-time, 
and a factor of $4$ in the denominator adjusts for taste degrees of freedom. 
Its disconnected susceptibility is given by
\begin{align}
\chi_\text{disc} = \frac{1}{16 N_{\sigma}^3 N_{\tau}}
\left[
\av{ \left( \Tr\frac{1}{D_{-1/3}+m} \right)^2 }
-
\av{ \Tr\frac{1}{D_{-1/3}+m}  }^2
\right].
\end{align}

While chiral condensate and its susceptibility are related to chiral symmetry we also compute the Polyakov loop which is related to the deconfinement transition in a pure glue system
\begin{equation}
P = \frac{1}{V}\left\langle \sum_{\vec{x}}\mathrm{Tr}\prod_{t}U_4(t,\vec{x})\right\rangle,
\end{equation}
and its susceptibility $\chi_{\rm P}$.

We will also analyse the Binder cumulants $B_M$ of order parameters $M$, e.g. the chiral 
condensate\cite{binder1981critical} and the Polyakov loop as well as the constructed order parameter from a mixture of the chiral condensate and the gauge action (cf. Eq.~\ref{eq:order_parameter}). $B_{M}$ is defined 
as follows
\begin{align}
B_{M} (\beta, N_b) = \frac{\langle (\delta M)^4 \rangle} {\langle(\delta M)^2 \rangle^2}\; ,
%B_4 (\beta, N_b) = \frac{ \av {(\deltapsi)^4} }{ \av{(\deltapsi)^2}^2 },
\end{align}
where $\delta M=M-\langle M \rangle$
gives the deviation of $M$ from its mean value
on a given gauge field configuration.
From different distributions of $M$ in different phases the value 
of $B_M$ can be obtained and used to distinguish phase 
transitions.
Given that the chiral condensate is the order parameter of the phase transition~\footnote{The same holds true for other order parameters.}, for a first order phase transition, $B_{\bar{\psi}\psi}=1$~\cite{Ejiri:2007ga}; for a crossover $B_{\bar{\psi}\psi}\simeq3$~\cite{Ejiri:2007ga}; for a second order transition belonging to a Z(2) universality class, $B_{\bar{\psi}\psi}\simeq1.6$~\cite{Blote:1995zik}.

Since our lattice is rather coarse and simulated systems are in the proximity of transitions, we also estimate the effective number of independent configurations $N_\text{eff}$ given by
\begin{align}
N_\text{eff} = \frac{\# \text{ of trajectories} }{ 2 \tau_\text{int} },
\end{align}
where $\tau_\text{int}$ is the integrated autocorrelation time~\footnote{
The autocorrelation time and its error are given in Appendix \ref{appendix:acc} where $\tau_\text{int}$ are rounded in integer for simplicity.
} for the chiral condensate. Obtained results of $\tau_\text{int}$ are listed in Appendix \ref{appendix:acc}. Hereafter we will show our results of various quantities in a dimensionless way, i.e. all in units of lattice spacing $a$.

\section{Results} \label{sec:results}
In the first subsection, we discuss the observables obtained from a fixed volume $N_\sigma = 16$ and the history of the chiral condensate at vanishing and nonzero values of $\hat{b}$.
In the second subsection, we study the volume dependences of chiral observables and the Polyakov loop as well as their susceptibilities at $\hat{b} = 0$ and $1.5$ obtained from lattices with $N_\sigma$=8, 16, 20 and 24.
In the third subsection, we show results based on a appropriate order parameter constructed from a combination of the chiral condensate and the gluon action.

\subsection{$a \sqrt{eB}$ dependence of observables obtained on $N_\sigma=16$ lattices}

\begin{figure}[htbp!]
	\begin{center}
				\includegraphics[width=0.46\textwidth]{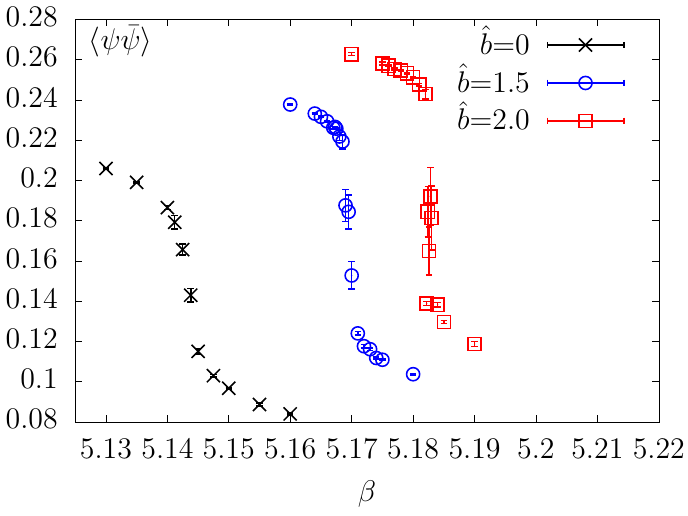}
				\includegraphics[width=0.46\textwidth]{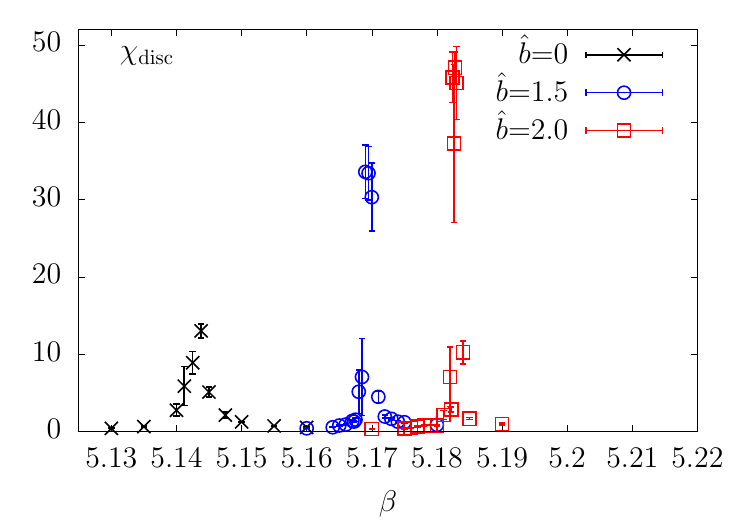}
	\end{center}
	\caption{
		The chiral condensate (left) and disconnected chiral susceptibility (right) at $\hat{b}=0,$ 1.5 and 2 obtained on lattices with $N_\sigma=16$ as a function of $\beta$.
		\label{fig:pbp_L16} }
\end{figure}

We show the chiral condensate and its disconnected susceptibility for $\hat{b} = 0, 1.5$ and 2 obtained from lattices with $N_\sigma$=16  as a function of the inverse gauge coupling $\beta$ in the left and right plot of \fig\ref{fig:pbp_L16}, respectively. We recall that the temperature is an increasing function of $\beta$. 
As seen from the left plot the value of the chiral condensate is enhanced by the magnetic field in the whole temperature region. Namely only the magnetic catalysis is observed. One can also see that the critical $\beta$ value where the chiral condensate drops most rapidly becomes larger as the magnetic field strength increases. This indicates that the transition temperature increases as the magnetic field becomes stronger. Meanwhile as the strength of the magnetic field increases the dropping of chiral condensates becomes more rapidly. This means that the transition becomes stronger with stronger magnetic fields. These two observations are also visible in the behaviour of disconnected chiral susceptibilities as shown in the right plot of \fig\ref{fig:pbp_L16}. I.e. the peak location of disconnected chiral susceptibility shifts to a larger value of $\beta$ and the peak height of disconnected chiral susceptibility increases as the strength of magnetic field increases.

\begin{figure}[htbp!]
	\begin{center}
		\includegraphics[width=0.44\textwidth]{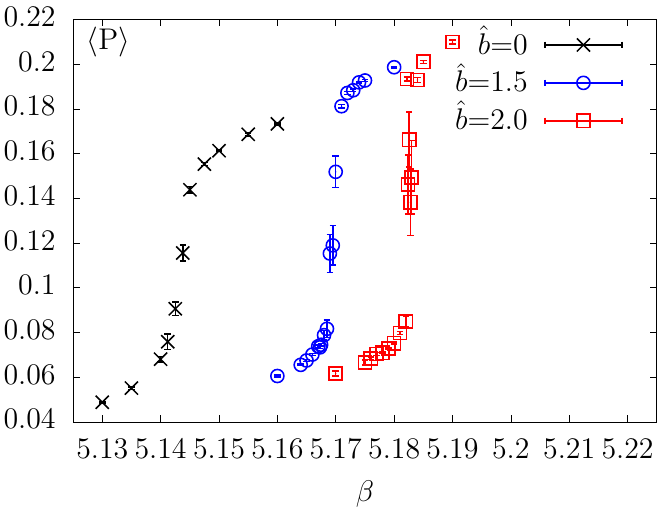}
		~\includegraphics[width=0.46\textwidth]{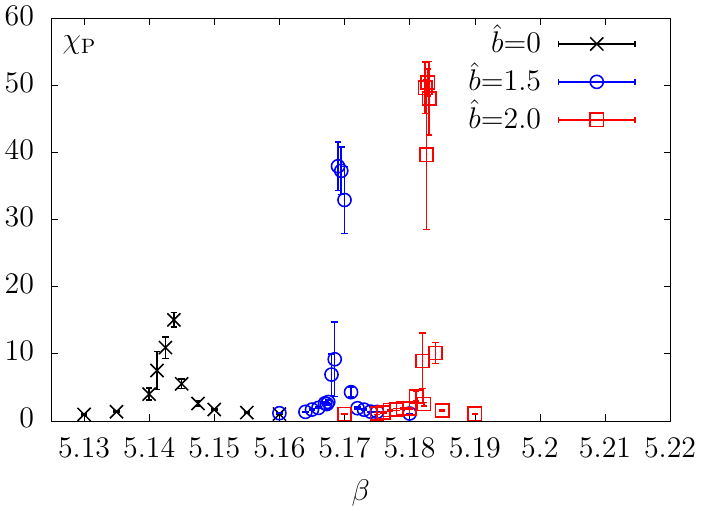}
	\end{center}
	\caption{
		Same as \fig\ref{fig:pbp_L16} but for the Polyakov loop and its susceptibility.
		\label{fig:pol_L16} }
\end{figure}

We show similar plots as \fig\ref{fig:pbp_L16} for the Polyakov loop and its susceptibility at $\hat{b} = 0, 1.5$ and 2 in \fig\ref{fig:pol_L16}. The peak location of the Polyakov loop susceptibility as well as the inflection point of the Polyakov loop shifts to larger values of $\beta$, i.e. higher transition temperature with stronger magnetic field. This is similar to the observation from chiral observables. Besides that transition temperatures signalled by the Polyakov loop and its susceptibility are close to those obtained from chiral condensates and disconnected chiral susceptibility. 
What's more, it is interesting to see that the peak height of the Polyakov loop susceptibility also becomes higher in a stronger magnetic field, although the Polyakov loop is an order parameter for the confinement-deconfinement phase transition of a pure glue system while the chiral condensate is the order parameter of QCD transitions with vanishing quark massless. This could be due to the fact that neither of these two quantities are the true order parameter but a part of the true order parameter in $N_f=3$ QCD\footnote{Note that the peak height of Polyakov loop susceptibility also increases as the system approaches the first order phase transition region with smaller values of quark masses for the case of zero magnetic field strength $\hat{b}=0$~\cite{Karsch:2001nf}.}~\cite{Karsch:2001nf,Bazavov:2017xul}.

\begin{figure}[htbp!]
	\begin{center}
		\includegraphics[width=0.44\textwidth]{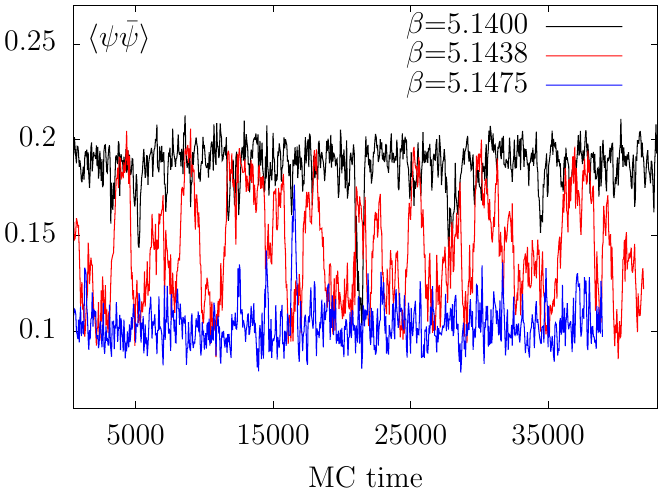}
		\includegraphics[width=0.44\textwidth]{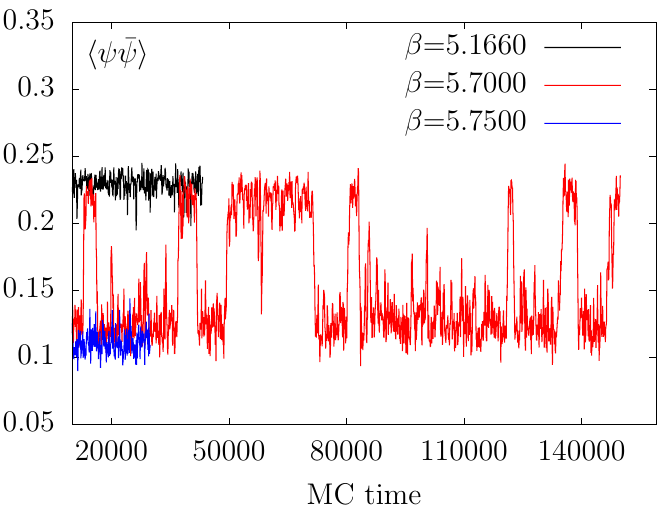}
		\includegraphics[width=0.44\textwidth]{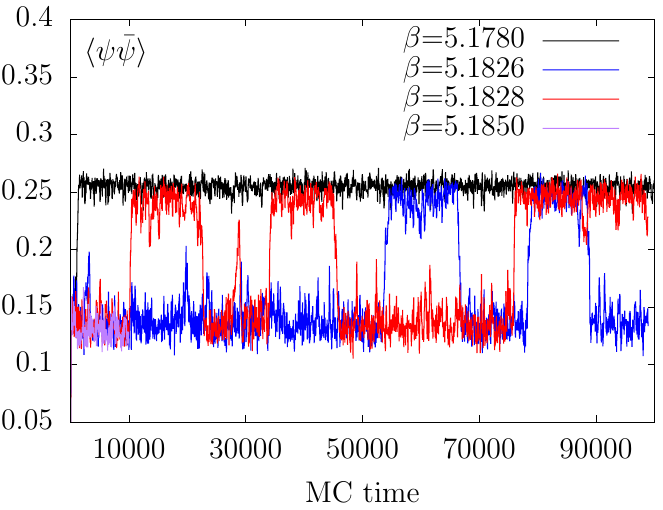}
		\includegraphics[width=0.44\textwidth]{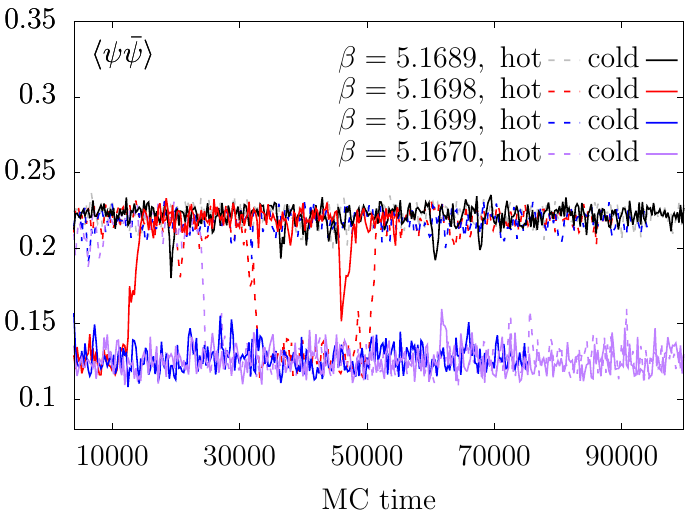}
	\end{center}
	\caption{
		Monte Carlo time history of chiral condensates. The x axes are in units of trajectories of molecular dynamics. Top left: with $\hat{b}=0$ and $N_\sigma=16$. Top right: with $\hat{b}=1.5$ and $N_\sigma=16$. Bottom left: with $\hat{b}=2$ and $N_\sigma=16$. Bottom right: with $\hat{b}=1.5$ and $N_\sigma=24$. 
		\label{fig:pbp_history_L16} }
\end{figure}

At vanishing magnetic field the transition is a crossover, and as the magnetic field strength becomes larger the QCD transition becomes stronger. In particular the jumping behavior of the chiral condensate and the Polyakov loop at $\hat{b}$=2 is seen from left plots of Fig.~\ref{fig:pbp_L16} and Fig.~\ref{fig:pol_L16}. This might indicate a first order phase transition. We investigate the nature of the transition by further looking into the time history of chiral condensates. We show in Fig.~\ref{fig:pbp_history_L16} the time history of chiral condensates near the transition temperature at $\hat{b}=0$ (Top left), $\hat{b}=1.5$ (Top right) and $\hat{b}=2$ (Bottom left) on lattices with $N_\sigma=16$. As expected that at vanishing magnetic field ($\hat{b}$=0) there does not exist any metastable behavior, while in the case of $\hat{b}=$1.5 and 2 the metastable behavior becomes obvious. To confirm the metastable behavior seen from the volume of $N_\sigma$=16, we also study the case of $\hat{b}=$1.5 with a larger volume, i.e. on $24^3\times4$ lattices. Since in the first order phase transition the tunnelling rate between two metastable states becomes smaller in the larger volume, here we rather investigate on the time history of the chiral condensate obtained from two different kinds of streams, i.e. one starting from a unit configuration (cold) and the other one starting from a random configuration (hot). If there is no first order phase transition chiral condensates obtained from the cold and hot starts will always overlap after thermalization, and if there exist a first order phase transition the two steams from cold and hot starts will stay apart and tunnel from one to the other as the two metastable states in the first order phase transition. The former is observed for the case of $\beta=5.1689$ while the latter is clearly seen for $\beta=5.1698$.

In a short summary, above observations suggest that the transition tends to be a first order for magnetic fields of $\hat{b}\geq$1.5.

% = = = = = = = = = = = = = = = = = = = = = = = = = = = = = = = = = = = = = = =
\subsection{Volume dependences of observables with $\hat{b}=0$ and $1.5$}

\begin{figure}[htbp!]
	\begin{center}
		\includegraphics[width=0.44\textwidth]{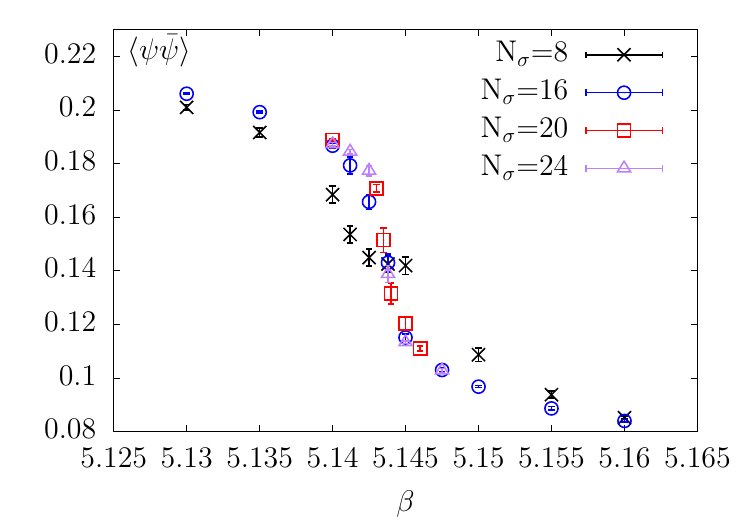}
		\includegraphics[width=0.44\textwidth]{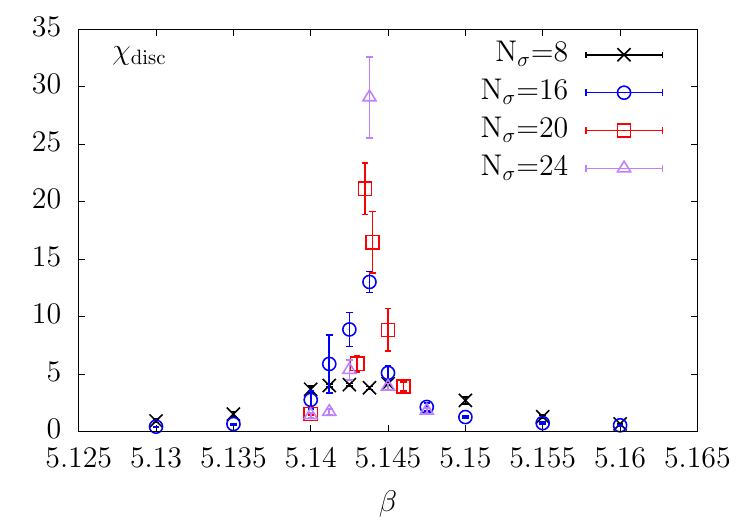}
		\includegraphics[width=0.44\textwidth]{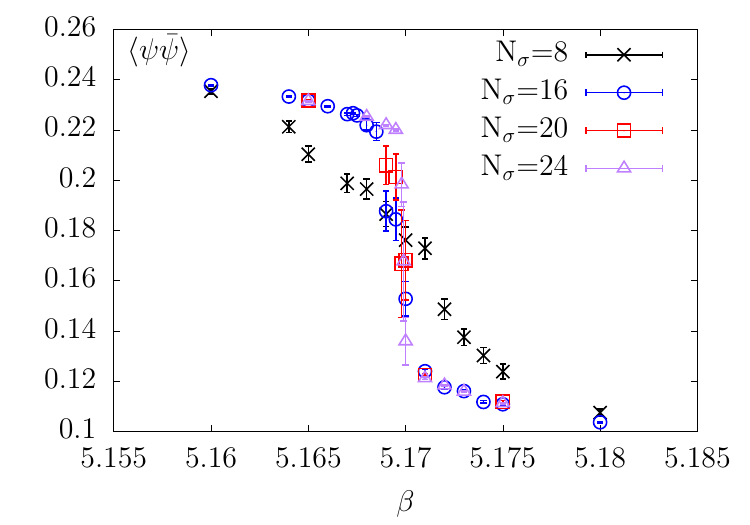}
		\includegraphics[width=0.44\textwidth]{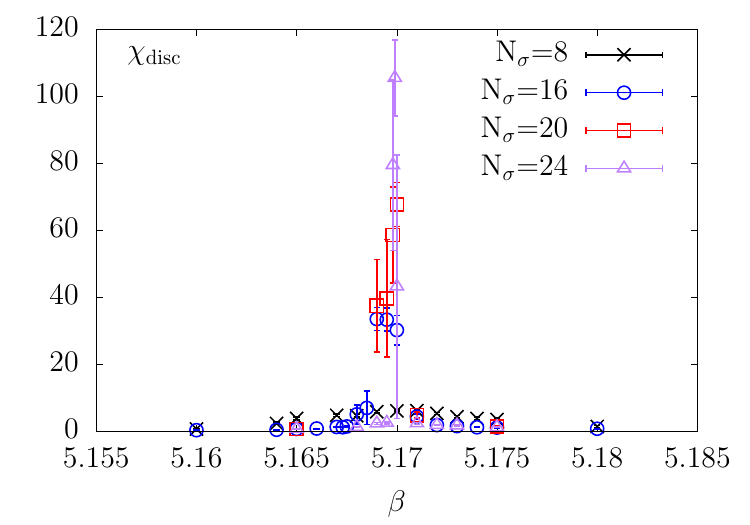}
	\end{center}
	\caption{
		Volume dependences of chiral condensates (left column) and its disconnected susceptibilities (right column) at vanishing magnetic field (top two plots) and $\hat{b}=$1.5 (bottom two plots). 
		\label{fig:pbp_av_b} }
\end{figure}
In this subsection, we discuss more on the volume dependence of
observables at $\hat{b}=0$ and $1.5$ to further confirm the onset of the first order phase transition at $\hat{b}\geq1.5$. In Fig.~\ref{fig:pbp_av_b} we show chiral condensates and their disconnected susceptibilities as a function of $\beta$ obtained from lattices with four different spatial sizes at $\hat{b}=0$ (Top two plots) and $\hat{b}$=1.5 (Bottom two plots). At vanishing magnetic field chiral condensates obtained from $N_\sigma=16,$ 20 and 24 are consistent among each other, while large deviations are seen for the results obtained from $N_\sigma$=8. 
In the case of $\hat{b}$=1.5 the finite size effect seems to appear at $\beta$ smaller and close to $\beta_c$ starting with a larger volume, i.e. $N_\sigma$=16. This could be due to the fact that the system with the presence of a magnetic field tends to have a stronger transition and consequently more statistics is needed to get robust results~\footnote{Although we generated two times more configurations at $\hat{b}$=1.5, the effective configurations, however, is two times less than that at $\hat{b}$=0.}, and that the correlation length in the system becomes longer in the proximity of the true phase transition. Nevertheless, the point where chiral condensates drops most rapidly are consistent among various volumes for both vanishing and nonzero magnetic fields. The is also reflected in the peak locations of disconnected susceptibilities shown in the right column of Fig.~\ref{fig:pbp_av_b}. In the case of a first order phase transition the disconnected chiral susceptibility should grow linearly in volume. It is more or less the case for disconnected susceptibilities at $\hat{b}$=1.5 as seen from bottom right plot of Fig.~\ref{fig:pbp_av_b} and Table~\ref{tab:betac}. And as expected that in the case of vanishing magnetic field the disconnected susceptibility grows slower than linearly in volume as there exists a crossover transition.

\begin{figure}[htbp]
	\begin{center}
		%\begin{tabular}{c}
		\includegraphics[width=0.46\textwidth]{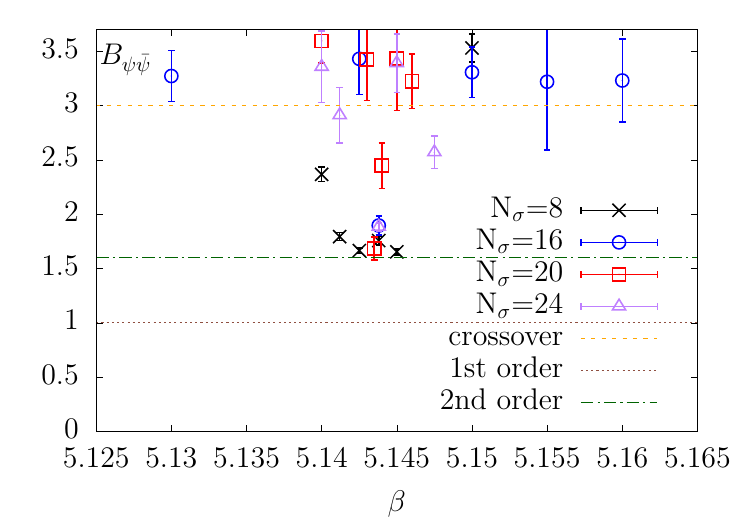}~
		\includegraphics[width=0.44\textwidth]{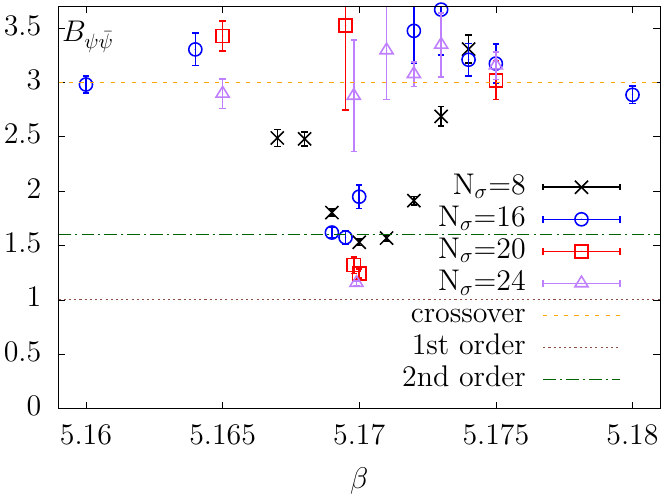}
	\end{center}
	\caption{
		Binder cumulants of chiral condensate at zero magnetic field (left) and nonzero magnetic field of $\hat{b}=$1.5 (right).
		\label{fig:binder_pbp_b} }
\end{figure}

We show Binder cumulants of chiral condensates at $\hat{b}=0$ and 1.5 in the left and right plot of Fig.~\ref{fig:binder_pbp_b}, respectively. One can see that at the critical $\beta$ where $\chi_{\rm disc}$ peaks  the Binder cumulant reaches to its minimum, which is almost independent of volume at $\hat{b}=0$ and only starts to saturate with $N_\sigma\geq$ 20 at $\hat{b}$=1.5. In nonzero magnetic fields the minimum values of Binder cumulant become smaller, i.e. shifting from about 1.6 in the vanishing magnetic field to about 1. This indicates that the transition becomes stronger and tends to become a  first order phase transition region at $\hat{b}$=1.5. This is consistent with what we found from chiral condensates and its susceptibilities. 

\begin{figure}[htbp!]
        \begin{center}
        \includegraphics[width=0.44\textwidth]{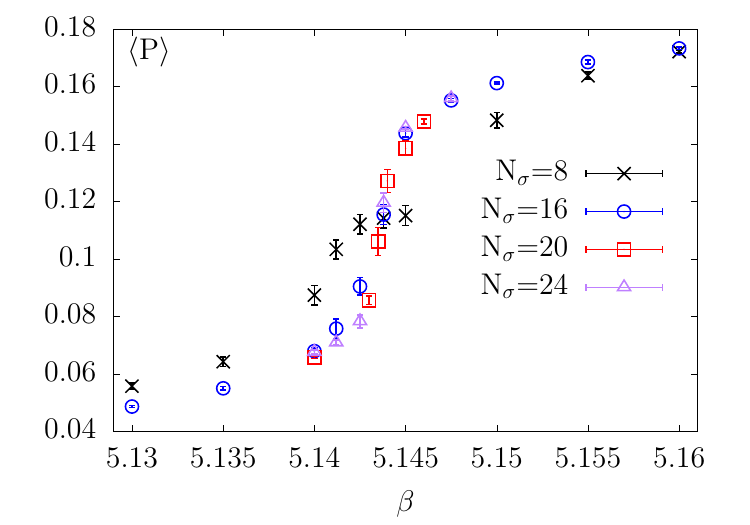}
        \includegraphics[width=0.44\textwidth]{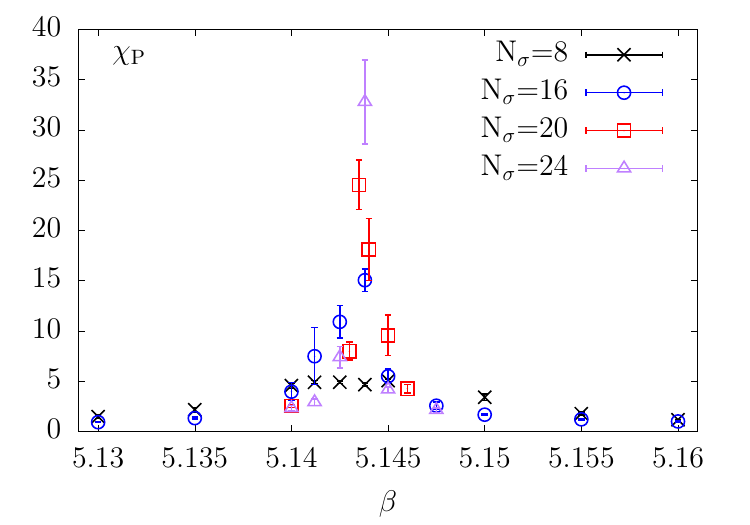}
         \includegraphics[width=0.44\textwidth]{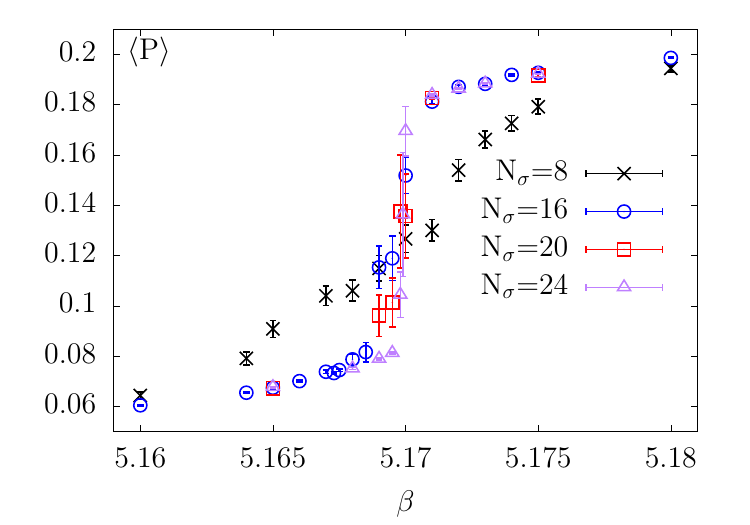}
        \includegraphics[width=0.44\textwidth]{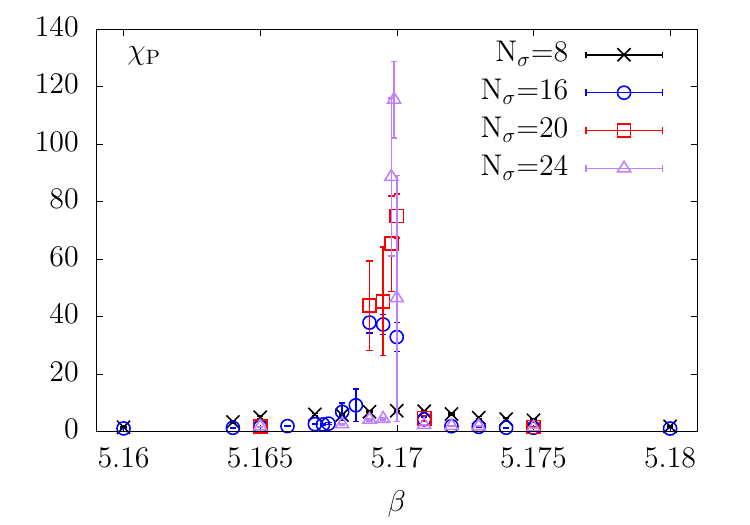}
        \end{center}
\caption{
Same as Fig.~\ref{fig:pbp_av_b} but for the Polyakov loop and its susceptibility.
\label{fig:pol_av_b} }
\end{figure}

\begin{figure}[htbp!]
	\begin{center}
		%\begin{tabular}{c}
		\includegraphics[width=0.44\textwidth]{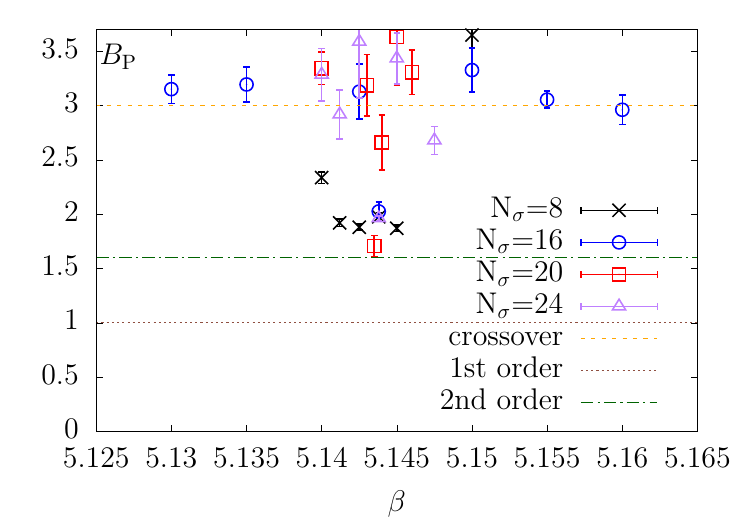}
				\includegraphics[width=0.44\textwidth]{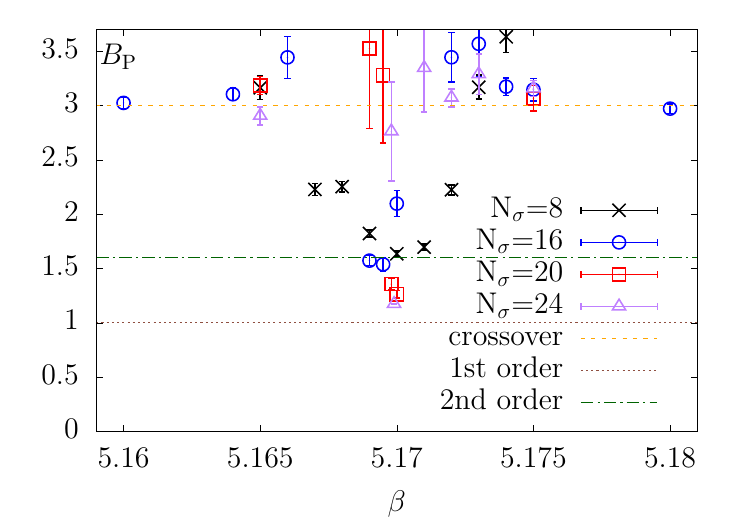}
	\end{center}
	\caption{
Same as Fig.~\ref{fig:binder_pbp_b} but for the Polyakov loop.
		\label{fig:binder_pol_b} }
\end{figure}

We show similar figures as Fig.~\ref{fig:pbp_av_b} and Fig.~\ref{fig:binder_pbp_b} for Polyakov loops in Fig.~\ref{fig:pol_av_b} and Fig.~\ref{fig:binder_pol_b}. The observation is similar to that from chiral observables. The volume dependences of Polyakov loops and corresponding susceptibilities are similar to those of chiral observables. Besides, the value for the Binder cumulant has the same tendency. The peak heights of susceptibilities and the minimum values of Binder cumulants for the chiral condensate and the Polyakov loop are summarized in \tab \ref{tab:betac}.

\begin{table}[htb]
	\begin{tabular}{c|c||c|c|c|c|c} 
		$N_\sigma$ & $a\sqrt{eB}$ & $\beta_\text{c}$ & $\chi_\text{disc} (\beta_\text{c}) $& $B_{\psi\bar{\psi}} (\beta_\text{c}) $  
		& $\chi_\text{P} (\beta_\text{c}) $& $B_\text{P} (\beta_\text{c}) $\\ \hline \hline
		8 & 0 & 5.1450  & 4.1(1) & 1.65(3) & 5.1(1)   & 1.87(3)\\ 
		16 & 0 & 5.1438 & 13.0(9) & 1.90(9) & 15(1) & 2.03(9)\\ 
		20 & 0 & 5.1435 & 21(2)  & 1.7(1) & 25(2)     & 1.7(1)\\ 
		24 & 0 & 5.1438 & 29(4) & 1.88(4) & 33(4)    & 1.96(2) \\ \hline 
		8 & 1.5& 5.1710 &  6.2(2) & 1.57(3) & 7.1(2) & 1.70(3) \\ 
		16 & 1.5 & 5.1690 & 34(3)  & 1.62(5) & 38(4) & 1.57(4) \\ 
		20 & 1.5 & 5.1700 & 68(7)  & 1.24(4) & 75(8) & 1.26(4)\\ 
		24 & 1.5 & 5.1699 & 105(11) & 1.15(2) & 115(13) &  1.173(3)
	\end{tabular}
	\caption{Values of disconnected chiral susceptibility $\chi_{\rm disc}$, Polyakov loop susceptibility $\chi_{\rm P}$, and the Binder cumulants for the chiral condensate and the Polyakov loop at $\hat{b}=0$ and 1.5. $\beta_\text{c}$ is the value of $\beta$ where $\chi_\text{disc} (\beta) $ peaks.
		\label{tab:betac}   
	}
\end{table}

\subsection{Binder cumulants and disconnected susceptibilities of the order parameter}

In the vicinity of a critical point in 3-flavor QCD, the chiral condensate itself is not a true order parameter, but is part of a mixture of operators that defines the true order parameter $M$~\cite{Karsch:2001nf,Bazavov:2017xul}. In three flavor QCD, such an order parameter can be constructed as a combination of the plaquette action and the chiral condensate as follows~\cite{Karsch:2001nf,Bazavov:2017xul} 
\begin{align}
M(\beta,s) = {\bar{\psi}\psi}(\beta) + s \,\frac{1}{N_\sigma^3 N_\tau} S_\text{gauge}  (\beta) ,
\label{eq:order_parameter}
\end{align}
where $S_\text{gauge} =6 N_\sigma^3 N_\tau \tilde{P}$ and $\tilde{P}$ is the plaquette.
Correspondingly, its susceptibility is,
\begin{align}
\chi_\text{mixed} = \frac{1}{16 N_{\sigma}^3 N_{\tau}}
\left[
\av{ \left( M(\beta,s) \right)^2 }
-
\av{M(\beta,s)  }^2
\right].
\end{align}
In this work, we do not intend to determine the mixing parameter $s$ and just use the value obtained in the previous work for $a\sqrt{eB}=0$ in \cite{Smith:2011pm}, i.e. $s=-0.8$. As will be seen next the results obtained using $s=-0.8$ does not change from $s=0$ qualitatively (cf. Table~\ref{tab:betac}).

\begin{figure}
		\includegraphics[width=0.48\textwidth ]{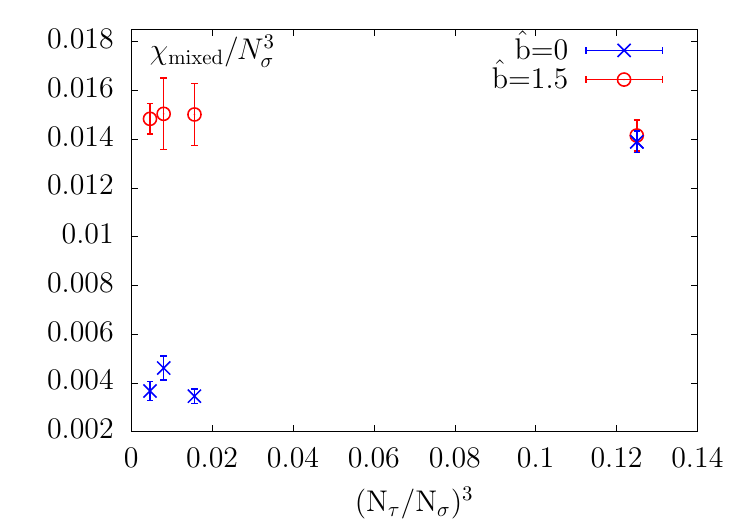}
\includegraphics[width=0.48\textwidth]{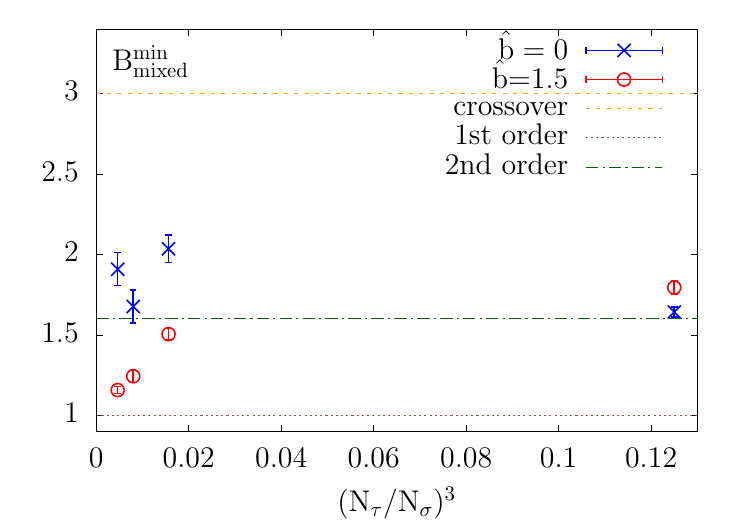}
\caption{Left: $\chi_\text{mixed}$ divided by spatial volume (left) and Binder cumulant of $M$ at $\beta=\beta_c$ obtained with $s=-0.8$ at $\hat{b}=0$ and 1.5 as a function of the inverse spatial volume represented by $(N_\tau/N_\sigma)^3$.
	\label{fig:M} }
\end{figure}

We show in the left panel of Fig.~\ref{fig:M} the order parameter susceptibility $\chi_\text{mixed}$ divided by the spatial volume at $\hat{b}=0$ and 1.5.
If the phase transition is of a first order $\chi_\text{mixed}$ divided by the spatial volume should be a constant in volume. It is clearly seen that at $\hat{b}$=1.5 data points obtained from lattices with different volumes all overlap at about 0.015 which holds true also towards the infinite volume limit of $(N_\tau/N_\sigma)^3\rightarrow0$, while it is not the case for $\hat{b}=0$.
The Binder cumulant of the order parameter is shown in the right panel of \fig\ref{fig:M}.
Data points for $\hat{b}=0$ are close the $\mathbb{Z}_2$ line, so it seems to belong to a weak second order or crossover transition in the thermodynamic limit. On the other hand, data points for $\hat{b}= 1.5$ approach 1 towards the infinite volume limit of $(N_\tau/N_\sigma)^3\rightarrow0$. This is again consistent with the behavior in the first-order phase transition.

In summary, we conclude that the system with $am = 0.03$ with magnetic fields $a\sqrt{eB}\geq$1.5 ($eB \gtrsim 11 m_\pi^2$) exists a first order phase transition.

\section{Conclusion} \label{sec:conclusion}

We have investigated the chiral phase structure of three flavor QCD in the magnetic field. 
The simulations are performed with 3-mass degenerate flavors of standard staggered quark and the Wilson plaquette gauge action on $N_\tau=4$ lattices. We started with simulations at a fixed quark mass of $am=0.03$ at vanishing magnetic field where a crossover transition is observed. After turning on the magnetic field we studied dependences of chiral observables, i.e. chiral condensates and corresponding susceptibilities as well as Polyakov loops on the magnetic field strength. We found that chiral condensates increase with the magnetic field strength, namely magnetic catalyses in the whole temperature window. And transition temperatures determined from both chiral condensates and Ployakov loops as well as their susceptibilities increase with increasing magnetic field strength. From the dropping behavior of chiral condensates and Polyakov loops near the transition temperature and the peak heights of susceptibilities we found that the strength of transition increases with increasing magnetic field strength. We thus checked the time history of the chiral condensate, and the metastable behavior of chiral condensates is observed at $a\sqrt{eB}\geq 1.5$. This indicates a first order phase transition occurring to systems at $a\sqrt{eB}\geq 1.5$. We further investigate a more appropriate order parameter which is constructed from the chiral condensate and the plaquette gauge action. We studied the volume dependence of order parameter susceptibility and the Binder cumulant of the order parameter, and confirmed that there exists a first order phase transition at $a\sqrt{eB}\geq 1.5$. Based on the earlier estimate we find that $a\sqrt{eB}\geq 1.5$ corresponds to $eB/m_\pi^2\gtrsim 11$ with the pion mass at zero magnetic field $m_\pi\sim280$ MeV.
 
We do not find any signs of inverse magnetic catalysis and the reduction of transition temperature as a function of the magnetic field strength, and this is consistent with the results obtained from lattice  studies of  $N_f=2$ QCD with the standard staggered quarks and larger-than-physical pions \cite{DElia:2010abb,d2011chiral}. Since our findings of magnetic catalysis and increasing of transition temperature with the magnetic field strength even holds in the regime occurring a first order phase transition this suggests that the discrepancy from results using the improved staggered fermions with physical pions is mainly due to the lattice cutoff effects. 

It is worth recalling that in the case of lattice studies using improved staggered fermions the strength of transition also increases with increasing magnetic field strength and a first order phase transition has not yet been observed in such studies. Although we find a first order phase transition in the current study using the standard staggered fermions, we do not intend to provide a precise determination of a critical magnetic field in which the transition turns into a first/second order phase transition in the current discretization scheme.  We will leave it for future studies using improved staggered fermions, such as HISQ fermions to achieve more realistic results on the chiral phase structure of $N_f=3$ QCD~\cite{Tomiya:2019nym}. As in current studies severe critical slowing down is expected for simulations in the vicinity of phase transitions with larger volumes (see Appendix~\ref{appendix:acc}), the autocorrelation length will probably become even longer in simulations with smaller lattice spacings or improved actions, in which the multicanonical method \cite{Berg:1992qua,Bonati:2018blm} or other extended Monte Carlo method reviewed in \cite{Iba:2000qb} might help.

\section*{Acknowledgement}
We thank Swagato Mukherjee for the early involvement of the work as well as enlightening discussions. This work was supported  by the NSFC under grant no. 11535012 and no. 11775096,  RIKEN Special Postdoctoral Researcher program, Deutsche Forschungsgemeinschaft project number 315477589 -- TRR 211 and European Union H2020-MSCA-ITN-2018-813942 (EuroPLEx).
The numerical simulations have been performed on the GPU cluster in the Nuclear Science Computing Center at Central China Normal University (NSC$^3$).

\appendix
\newcommand{\nconf}{ {N_\text{trj}} }
\section{Autocorrelation time} \label{appendix:acc}
Here we explain the autocorrelation time to make this paper self-contained.
A sequence of configurations are affected by the autocorrelation, which is evaluated by the autocorrelation function.
However, the autocorrelation function itself is a statistical object,
so we cannot determine it exactly. Instead we calculate the approximated autocorrelation function
\cite{Wolff:2003sm, Madras:1988ei} defined by,
\begin{align}
\bar\Gamma(\tau) = \frac{1}{\nconf - \tau} \sum_{c}^\nconf (O_c-\bar{O})(O_{c+\tau}-\bar{O}),
\label{autocorrelation_function}
\end{align}
where $O_c = O[ U^{(c)} ]$ is the value of operator $O$ for the $c$-th configuration $U^{(c)}$ and $\tau$ is the fictitious time of HMC.
$\nconf$ is the number of trajectories.
Conventionally, the normalized  autocorrelation function $\bar\rho(\tau)=\bar\Gamma(\tau) /\bar\Gamma(0) $
is used.

The integrated autocorrelation time $\tau_\text{int}$ approximately quantifies 
effects of autocorrelation. This is given by,% which is given by,
\begin{align}
\tau_\text{int} = \frac{1}{2} + \sum^{W}_{\tau=1} \bar\rho(\tau).
\label{tau_int}
\end{align}
We can regard two configurations separated by $2\tau_\text{int} $ as independent.
In practice, we determine a window size $W$ as a first point $W = \tau $, where $\bar{\Gamma}(\tau)<0$
for the smallest $\tau$.
The statistical error of integrated autocorrelation time
is estimated by the Madras--Sokal formula \cite{Madras:1988ei, Luscher:2005rx},
\begin{align}
\av{\delta \tau_\text{int}^2 }
\simeq
\frac{4W + 2}{\nconf} \tau_\text{int}^2, \label{eq:ac_time_error}
\end{align}
and we use the square root of \eqref{eq:ac_time_error} for an estimate of the error on the autocorrelation time,
 $\sqrt{\av{\delta \tau_\text{int}^2 }} \equiv \Delta \tau_\text{int}$.

\section{Critical slowing down}

\begin{figure}[htbp!]
	\begin{center}
		\includegraphics[width=0.8\textwidth]{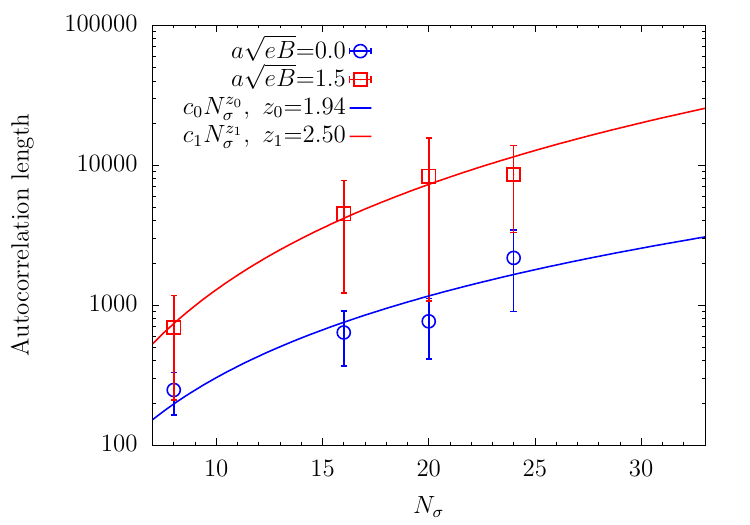}
	\end{center}
	\caption{
		The largest autocorrelation as a function of the spatial size $N_\sigma$ at 
		$\hat{b} \equiv a\sqrt{eB} = 0$ and 1.5.
		\label{fig:critical_slowing_down} }
\end{figure}
The estimated autocorrelation length for the chiral condensate and the number of independent configurations
are listed in \tab \ref{tab:stat_L16}
and \tab \ref{tab:stat_L24}. It can be observed that the autocorrelation time
becomes longer in the presence of a magnetic field. This is due to the fact that the system is close to the critical region as discussed in the main text.

In \fig\ref{fig:critical_slowing_down}, we estimate the dynamic critical exponent of the system by using
the longest autocorrelation length in each volume.
A fit ansatz of $\log \tau_\text{int}(N_\sigma) = z \log N_\sigma + c$ with fit parameters $z$ and $c$ is adopted and this fit ansatz is the same form as $\tau_\text{int}(N_\sigma) \sim N_\sigma^{z } $, which is the definition 
of the dynamic critical exponent.  One can see that the system is affected by severe critical slowing down at $\hat{b}\equiv\sqrt{a^2 eB}=1.5$. It is thus practically challenging to investigate
the first-order phase transition by using conventional direct simulations with HMC in the thermodynamic limit.

\section{Summary of statistics}
We summarize our simulation parameters and statistics
in \tab \ref{tab:stat_L8} - \ref{tab:stat_L24}. 
If the statistics is sufficient the binning size for the Jackknife analysis is taken as the $\text{Bin-size} \gtrsim 2\tau_\text{int}$, otherwise the $\text{Bin-size}$ is adjusted such that it is the divisor of number of trajectories by around 10. In the latter case, the error might be underestimated.

\begin{table}[!hptb]
\begin{tabular}{c|c|c|c|c|c|c||c} 
$N_\sigma$ & $\beta$ & $N_b$ & Bin-size & $\tau_\text{int}$ & $\Delta \tau_\text{int}$ & Traj. & $N_\text{eff}$ \\ \hline\hline
8 & 5.1300 & 00 & 500 & 73 & 12 & 49000 & 334 \\ 
8 & 5.1350 & 00 & 500 & 87 & 14 & 43000 & 246 \\ 
8 & 5.1400 & 00 & 500 & 189 & 49 & 46000 & 122 \\ 
8 & 5.1412 & 00 & 500 & 224 & 62 & 49500 & 110 \\ 
8 & 5.1425 & 00 & 500 & 198 & 54 & 49500 & 125 \\ 
8 & 5.1438 & 00 & 500 & 141 & 30 & 40500 & 144 \\ 
8 & 5.1450 & 00 & 500 & 194 & 48 & 49000 & 126 \\ 
8 & 5.1500 & 00 & 600 & 247 & 83 & 49200 & 100 \\ 
8 & 5.1550 & 00 & 500 & 145 & 55 & 49000 & 169 \\ 
8 & 5.1600 & 00 & 500 & 27 & 3 & 49000 & 920 \\ 
\end{tabular}
\begin{tabular}{c|c|c|c|c|c|c||c} 
$N_\sigma$ & $\beta$ & $N_b$ & Bin-size & $\tau_\text{int}$ & $\Delta \tau_\text{int}$ & Traj. & $N_\text{eff}$ \\ \hline\hline
8 & 5.1600 & 08 & 500 & 78 & 14 & 48500 & 310 \\ 
8 & 5.1640 & 08 & 500 & 159 & 34 & 50500 & 159 \\ 
8 & 5.1650 & 08 & 500 & 200 & 54 & 50500 & 126 \\ 
8 & 5.1670 & 08 & 500 & 275 & 72 & 50500 & 92 \\ 
8 & 5.1680 & 08 & 1000 & 233 & 58 & 50000 & 107 \\ 
8 & 5.1690 & 08 & 1000 & 279 & 72 & 50000 & 90 \\ 
8 & 5.1700 & 08 & 1500 & 690 & 481 & 49500 & 36 \\ 
8 & 5.1710 & 08 & 500 & 207 & 44 & 50500 & 122 \\ 
8 & 5.1720 & 08 & 600 & 301 & 86 & 50400 & 84 \\ 
8 & 5.1730 & 08 & 500 & 248 & 66 & 50500 & 102 \\ 
8 & 5.1740 & 08 & 500 & 202 & 52 & 50500 & 125 \\ 
8 & 5.1750 & 08 & 500 & 274 & 90 & 50500 & 92 \\ 
8 & 5.1800 & 08 & 500 & 76 & 13 & 50500 & 331 \\ 
\end{tabular}
  \caption{Statistics for $N_\sigma = 8$. Trajectories for thermalization are already discarded. Here one trajectory denotes one time unit in the molecular dynamics.}
   \label{tab:stat_L8}
\end{table}
 \begin{table}[!phtb]
\begin{tabular}{c|c|c|c|c|c|c||c} 
$N_\sigma$ & $\beta$ & $N_b$ & Bin-size & $\tau_\text{int}$ & $\Delta \tau_\text{int}$ & Traj. & $N_\text{eff}$ \\ \hline\hline
16 & 5.1300 & 00 & 500 & 24 & 4 & 9500 & 201 \\ 
16 & 5.1350 & 00 & 500 & 28 & 5 & 9500 & 172 \\ 
16 & 5.1400 & 00 & 500 & 193 & 46 & 41500 & 107 \\ 
16 & 5.1412 & 00 & 2000 & 636 & 270 & 38000 & 30 \\ 
16 & 5.1425 & 00 & 1000 & 447 & 150 & 41000 & 46 \\ 
16 & 5.1438 & 00 & 1000 & 396 & 117 & 40000 & 50 \\ 
16 & 5.1450 & 00 & 500 & 190 & 47 & 40500 & 107 \\ 
16 & 5.1475 & 00 & 500 & 97 & 19 & 37500 & 193 \\ 
16 & 5.1500 & 00 & 500 & 42 & 6 & 38500 & 453 \\ 
16 & 5.1550 & 00 & 500 & 25 & 5 & 7000 & 142 \\ 
16 & 5.1600 & 00 & 500 & 22 & 5 & 7000 & 163 \\ 
\end{tabular}
\begin{tabular}{c|c|c|c|c|c|c||c} 
$N_\sigma$ & $\beta$ & $N_b$ & Bin-size & $\tau_\text{int}$ & $\Delta \tau_\text{int}$ & Traj. & $N_\text{eff}$ \\ \hline\hline
16 & 5.1600 & 32 & 500 & 40 & 8 & 29500 & 373 \\ 
16 & 5.1640 & 32 & 500 & 36 & 4 & 47500 & 659 \\ 
16 & 5.1650 & 32 & 500 & 60 & 12 & 36000 & 298 \\ 
16 & 5.1660 & 32 & 500 & 55 & 8 & 41000 & 375 \\ 
16 & 5.1670 & 32 & 500 & 100 & 25 & 42500 & 212 \\ 
16 & 5.1673 & 32 & 500 & 94 & 16 & 97500 & 518 \\ 
16 & 5.1675 & 32 & 500 & 136 & 21 & 97500 & 358 \\ 
16 & 5.1680 & 32 & 2000 & 630 & 169 & 102000 & 81 \\ 
16 & 5.1685 & 32 & 4000 & 1600 & 722 & 96000 & 30 \\ 
\end{tabular}
\begin{tabular}{c|c|c|c|c|c|c||c} 
$N_\sigma$ & $\beta$ & $N_b$ & Bin-size & $\tau_\text{int}$ & $\Delta \tau_\text{int}$ & Traj. & $N_\text{eff}$ \\ \hline\hline
16 & 5.1690 & 32 & 7000 & 3635 & 1925 & 140000 & 19 \\ 
16 & 5.1695 & 32 & 3000 & 1406 & 635 & 57000 & 20 \\ 
16 & 5.1700 & 32 & 8000 & 4481 & 3263 & 144000 & 16 \\ 
16 & 5.1710 & 32 & 500 & 265 & 56 & 97500 & 184 \\ 
16 & 5.1720 & 32 & 500 & 92 & 26 & 27500 & 150 \\ 
16 & 5.1730 & 32 & 500 & 63 & 13 & 27500 & 219 \\ 
16 & 5.1740 & 32 & 500 & 60 & 17 & 27500 & 230 \\ 
16 & 5.1750 & 32 & 500 & 38 & 6 & 27500 & 360 \\ 
16 & 5.1800 & 32 & 500 & 55 & 16 & 27500 & 249 
\end{tabular}
\begin{tabular}{c|c|c|c|c|c|c||c} 
$N_\sigma$ & $\beta$ & $N_b$ & Bin-size & $\tau_\text{int}$ & $\Delta \tau_\text{int}$ & Traj. & $N_\text{eff}$ \\ \hline\hline
16 & 5.1700 & 56 & 100 & 21 & 8 & 1500 & 36 \\ 
16 & 5.1750 & 56 & 100 & 26 & 13 & 1500 & 29 \\ 
16 & 5.1760 & 56 & 100 & 32 & 7 & 9000 & 140 \\ 
16 & 5.1770 & 56 & 100 & 47 & 12 & 9000 & 96 \\ 
16 & 5.1780 & 56 & 150 & 53 & 5 & 99000 & 939 \\ 
16 & 5.1790 & 56 & 150 & 70 & 9 & 81450 & 580 \\ 
16 & 5.1800 & 56 & 500 & 70 & 12 & 45500 & 325 \\ 
16 & 5.1810 & 56 & 500 & 195 & 35 & 95500 & 245 \\ 
16 & 5.1820 & 56 & 2500 & 1166 & 412 & 95000 & 41 \\ 
16 & 5.1822 & 56 & 2000 & 142 & 51 & 42000 & 148 \\ 
16 & 5.1824 & 56 & 7000 & 4641 & 3642 & 91000 & 10 \\ 
16 & 5.1826 & 56 & 8000 & 3890 & 2340 & 88000 & 11 \\ 
16 & 5.1828 & 56 & 9000 & 4820 & 3223 & 90000 & 9 \\ 
16 & 5.1830 & 56 & 8000 & 3923 & 2693 & 72000 & 9 \\ 
16 & 5.1840 & 56 & 100 & 865 & 393 & 49000 & 28 \\ 
16 & 5.1850 & 56 & 100 & 59 & 24 & 8900 & 75 \\ 
16 & 5.1900 & 56 & 100 & 23 & 8 & 1900 & 41 \\ 
\end{tabular}
  \caption{Same as Table~\ref{tab:stat_L8} but for $N_\sigma = 16$.}
   \label{tab:stat_L16}
\end{table}

\begin{table}[!hptb]
\begin{tabular}{c|c|c|c|c|c|c||c} 
$N_\sigma$ & $\beta$ & $N_b$ & Bin-size & $\tau_\text{int}$ & $\Delta \tau_\text{int}$ & Traj. & $N_\text{eff}$ \\ \hline\hline
20 & 5.1400 & 00 & 200 & 106 & 18 & 50800 & 240 \\ 
20 & 5.1430 & 00 & 500 & 237 & 68 & 29500 & 62 \\ 
20 & 5.1435 & 00 & 1500 & 764 & 353 & 30000 & 20 \\ 
20 & 5.1440 & 00 & 1300 & 682 & 358 & 29900 & 22 \\ 
20 & 5.1450 & 00 & 1000 & 537 & 244 & 30000 & 28 \\ 
20 & 5.1460 & 00 & 300 & 184 & 51 & 30600 & 83 \\ 
\end{tabular}
\begin{tabular}{c|c|c|c|c|c|c||c} 
$N_\sigma$ & $\beta$ & $N_b$ & Bin-size & $\tau_\text{int}$ & $\Delta \tau_\text{int}$ & Traj. & $N_\text{eff}$ \\ \hline\hline
20 & 5.1650 & 50 & 200 & 68 & 9 & 95200 & 696 \\ 
20 & 5.1690 & 50 & 5000 & 8327 & 7256 & 90000 & 5 \\ 
20 & 5.1695 & 50 & 3000 & 1750 & 1157 & 39000 & 11 \\ 
20 & 5.1698 & 50 & 7000 & 5680 & 6450 & 35000 & 3 \\ 
20 & 5.1700 & 50 & 5000 & 4919 & 4950 & 40000 & 4 \\ 
20 & 5.1710 & 50 & 150 & 291 & 78 & 66750 & 115 \\ 
20 & 5.1750 & 50 & 100 & 39 & 10 & 7800 & 99 
\end{tabular}
  \caption{Same as Table~\ref{tab:stat_L8} but for $N_\sigma = 20$.
  \label{tab:stat_L20}}
\end{table}

\begin{table}[!thpb]
\begin{tabular}{c|c|c|c|c|c|c||c} 
$N_\sigma$ & $\beta$ & $N_b$ & Bin-size & $\tau_\text{int}$ & $\Delta \tau_\text{int}$ & Traj. & $N_\text{eff}$ \\ \hline\hline
24 & 5.1400 & 00 & 500 & 168 & 129 & 7500 & 22 \\ 
24 & 5.1412 & 00 & 500 & 105 & 41 & 6500 & 31 \\ 
24 & 5.1425 & 00 & 1000 & 537 & 280 & 46900 & 44 \\ 
24 & 5.1438 & 00 & 4000 & 2171 & 1269 & 78000 & 18 \\ 
24 & 5.1450 & 00 & 500 & 176 & 38 & 74500 & 211 \\ 
24 & 5.1475 & 00 & 500 & 82 & 34 & 6500 & 40 \\ 
\end{tabular}
\begin{tabular}{c|c|c|c|c|c|c||c} 
$N_\sigma$ & $\beta$ & $N_b$ & Bin-size & $\tau_\text{int}$ & $\Delta \tau_\text{int}$ & Traj. & $N_\text{eff}$ \\ \hline\hline
24 & 5.1650 & 72 & 500 & 79 & 17 & 29500 & 187 \\ 
24 & 5.1680 & 72 & 500 & 102 & 17 & 47500 & 232 \\ 
24 & 5.1690 & 72 & 1000 & 215 & 28 & 180000 & 419 \\ 
24 & 5.1695 & 72 & 1500 & 174 & 26 & 102000 & 293 \\ 
24 & 5.1698 & 72 & 4000 & 4387 & 3022 & 76000 & 9 \\ 
24 & 5.1699 & 72 & 2000 & 2535 & 2917 & 14600 & 3 \\ 
24 & 5.1700 & 72 & 15000 & 8542 & 5236 & 186000 & 11 \\ 
24 & 5.1710 & 72 & 1000 & 111 & 25 & 42500 & 191 \\ 
24 & 5.1720 & 72 & 500 & 77 & 11 & 69500 & 453 \\ 
24 & 5.1730 & 72 & 500 & 111 & 29 & 51500 & 232 \\ 
24 & 5.1750 & 72 & 500 & 50 & 6 & 48000 & 481 \\ 
\end{tabular}
  \caption{Same as Table~\ref{tab:stat_L8} but for $N_\sigma = 24$.}
  \label{tab:stat_L24}
\end{table}

%\bibliography{ref}

\end{document}